\documentstyle[aps,12pt]{revtex}

 \def\beq{\begin{equation}}
 \def\eeq{\end{equation}}
 \def\beqa{\begin{eqnarray}}
 \def\eeqa{\end{eqnarray}}
 \def\nn{\nonumber}
\def\al{&&\!\!\!\!\!\!\!\!\!}
\def\a{&\!\!\!}
 \begin{document}
\begin{center}
 {\bf \large Black hole -- D-brane correspondence: An example}
 
 
 {Matteo Bertolini$^1$, Pietro Fre$^2$, Faheem Hussain$^3$, 
 Roberto Iengo$^1$, Carmen N\'u\~nez$^4$ and Claudio A.
 Scrucca$^1$}
 
$^1$ International School for Advanced Studies, Trieste, Italy\\
$^2$ Dipartimento di Fisica Teorica, Universita di Torino\\
 $^3$ International Centre for Theoretical Physics, Trieste, Italy\\
 $^4$ Instituto de Astronom\'{\i}a y F\'{\i}sica del Espacio (CONICET), 
 Buenos Aires,
 Argentina
 
 \vspace*{1cm}
 
\end{center} 
 
\begin{abstract}
{\bf Abstract:}
We explore the connection between D-branes and black holes in one particular
case: a $D3$-brane compactified to four dimensions on $T^6/Z_3$.
Using the  $D$-brane boundary state description we show the  
 equivalence with a double extremal N=2 black hole solution of four
dimensional supergravity.
 
 \end{abstract}
 
\section{Introduction}

The lack of a statistical mechanical theory of black
hole thermodynamics and the closely related problem of the black hole
information paradox are longstanding fundamental questions which can
now be precisely addressed. Explicit calculations are presently available due
to the recent progress in nonperturbative aspects of string theory
\cite{mal}.

The idea of relating black holes to elementary string states
is based on their common property of having a large 
degeneracy of states. However while the entropy of a Schwarszchild black hole
is proportional to the square of its mass, the 
logarithm of the degeneracy of elementary string states depends linearly on
the mass of the states. It was suggested that this discrepancy is due to
the large mass renormalization suffered by the string states due to quantum 
corrections, and thus could be avoided by 
 BPS states in superstring theories. Following the analogy, the BPS
condition on the states should correspond to the extremal condition on 
Reissner-Nordstr\"om black holes. 

A key step in the recent developments was the realization that in addition
to the states described by string fluctuations, there are also soliton states
in string theory, D-branes. 
The main advantage of using D-branes
instead of perturbative
 string states is that 
the event horizon of the corresponding black hole is non-singular and has
finite area. Thus the entropy for these black holes can be computed 
unambiguously, and can be compared with the corresponding microscopic
answer obtained from the counting of states of the D-brane. The two
calculations turn out to be in exact agreement, including the overall
numerical factor. Explicit calculations have been performed in many classes
of black holes which can be compared to different configurations of 
D-branes.   
This result was obtained initially
 for a five dimensional extremal black hole, and was later extended to
five dimensional rotating black holes, slightly non-extremal five
dimensional black holes, four dimensional extremal and slightly
non-extremal black holes. The five dimensional case was
considered first since one only needs three nonzero charges to obtain an
extremal black hole with regular horizon
in toroidal compactifications. In four dimensions one needs four
nonzero charges. For Calabi Yau compactifications not all the results of
the toroidal case hold. In particular, four different charges are no longer
needed in four dimensions. Another characteristic of Calabi Yau 
compactifications is that single D-brane black
holes are non singular. This is because the brane is wrapped on a topologically
non-trivial manifold and therefore can intersect itself, thus avoiding the
necessity of having
different branes in toroidal compactifications.
 
In this contribution we will explore the connection between D-branes and
black holes in one particular case. We will
explicity show how the analogy can be carried
through for a D3-brane compactified
to four dimensions on $T^6/Z_3$ by providing the evidence that supports
its identification  with a double
extremal N=2 black hole in four dimensions.
 In section 2 we summarize the boundary 
state description of a D3-brane wrapped on a 3-cycle of the $T^6/Z_3$
orbifold which was originally introduced in \cite{Hins1}. We also
recall the requirement
imposed by the BPS condition, namely that the cancellation of the force
between two identical D-branes in relative motion is
due to the exchange of the N=2 graviton multiplet containing the graviton and 
the graviphoton. 
This suggests that the
 classical solution corresponding to this configuration is a
Reissner-Nordstr\"{o}m black hole. 
In section 3 we introduce the four dimensional
 double extremal black hole solution of N=2
supergravity obtained by compactifying ten dimensional Type IIB 
supergravity on a Calabi Yau threefold. We also
show in this section how the correspondence between 
this solution and the D3-brane boundary state description can be established
\cite{bfis}.

\section{D3-branes on orbifolds}
 
 Let us consider a system of two D-branes in a type II
superstring theory 
 compactified down to four dimensions in the interesting case of the $Z_3$
orbifold, which breaks the 
supersymmetry down to N=2 (the branes further break it to N=1)
\cite{min,hin}.
This section is based on references \cite{Hins1,hin} where detailed
calculations will be
found.

The dynamics of these D-branes is determined by a one loop amplitude 
which can be conveniently evaluated in the boundary state formalism 
\cite{Polcai,Call}. 
In particular, one can compute the force between two D-branes moving with 
constant velocity, extending Bachas' result \cite{Bachas} to 
compactifications breaking some supersymmetry \cite{Hins1}.
This will be the key object  to establish the
D-brane-black hole correspondence. 
Analyzing the large distance behavior of the interaction and its velocity 
dependence, it is possible to read the charges with respect to the massless 
fields, and relate the various D-brane configurations to known solutions 
of the 4-dimensional low energy effective supergravity.

The amplitude for two D-branes moving with velocities $V_1 = \tanh v_1$, 
$V_2 = \tanh v_2$ (say along 1) and transverse positions $\vec Y_1$, 
$\vec Y_2$ (along 2,3), namely
\beq
\label{amp}
{\cal A}=\int_{0}^{\infty}dl \sum_s <B,V_1,\vec Y_1|e^{-lH}|B,V_2,\vec Y_2>_s
\eeq
is just a tree level propagation between the two boundary states which are 
defined to implement the boundary conditions specifying the branes.
The time is measured along the length of the cylinder $l$. 
There are two sectors, RR and NSNS, corresponding to periodicity and 
antiperiodicity of the fermionic fields around the cylinder, and after
the GSO projection there are four spin structures, R$\pm$ and NS$\pm$,
corresponding to all the possible periodicities of the fermions on the 
covering torus.

Let us consider a D-particle in four dimensional spacetime.
In the static case, the 0-brane has Neumann boundary
conditions in time and Dirichlet in
space. The velocity twists the 0-1 directions and gives them rotated boundary
conditions. 
The moving boundary state is most simply obtained by boosting the static one 
with a negative rapidity $v=v_1-v_2$ \cite{Billo}.
$$
|B,V,\vec Y> = e^{-ivJ^{01}}|B,\vec Y> \;.
$$
In the large distance limit $b \rightarrow \infty$ only world-sheets with
$l \rightarrow \infty$ will contribute, and momentum or winding in the 
compact directions can be safely neglected since they correspond to 
massive subleading components.

The moving boundary states
$$
|B,V_1,\vec Y_1>=\int\frac{d^{3}\vec k}{(2\pi)^{3}}
e^{i \vec k \cdot \vec Y_1} |B,V_1> \otimes|k_B> \;,\;\;
|B,V_2,\vec Y_2>=\int\frac{d^{3}\vec q}{(2\pi)^{3}}
e^{i \vec q \cdot \vec Y_2}|B,V_2> \otimes|q_B> \;,
$$
can carry only space-time momentum in the boosted combinations 
$k_B^\mu = (\sinh v_1 k^1, \cosh v_1 k^1, \vec k_T)$ and $q_B^\mu = 
(\sinh v_2 q^1, \cosh v_2 q^1, \vec q_T)$. 
Notice that because of their non zero velocity, the branes can also transfer
energy, and not only momentum as in the static case.

Integrating over the bosonic zero modes and taking into account momentum 
conservation ($k_B^\mu = q_B^\mu$), the amplitude factorizes into a bosonic 
and a fermionic piece:
\beq
{\cal A}=\frac 1{\sinh v} \int_{0}^{\infty}dl 
\int \frac {d^2 \vec k_T}{(2\pi)^2} e^{i \vec k \cdot \vec b} 
e^{-\frac {q_B^2}2} \sum_s Z_B Z^s_F 
=\frac 1{\sinh v} \int_{0}^{\infty} \frac {dl}{2\pi l} e^{- \frac {b^2}{2l}}
\sum_s Z_B Z^s_F \nn 
\eeq
with $Z_{B,F}=<B,V_1|e^{-lH}|B,V_2>^s_{B,F}$. From now on, 
$X^\mu \equiv X^\mu_{osc}$.

It is very convenient to group the fields into pairs,
\beqa
X^\pm = X^0 \pm X^1 \a \rightarrow \a ~ \alpha_{n},\beta_{n}= 
a^{0}_{n} \pm a^{1}_{n} \;, \nonumber \\
X^{i},X^{i*} = X^i \pm i X^{i+1} \a \rightarrow ~ \a 
\beta^i_{n},\beta^{i*}_{n}= a^{i}_{n}\pm i a^{i+1}_{n} \;,\;\; i=2,4,6,8  
\;, \nonumber \\
\chi^{A,B} = \psi^0 \pm \psi^1 \a \rightarrow ~ \a \chi^{A,B}_{n}= 
\psi^{0}_{n}\pm\psi^{1}_{n} \;,\nonumber \\
\chi^{i},\chi^{i*}= \psi^i \pm i \psi^{i+1} \a \rightarrow ~ \a \chi^{i}_{n},
\chi^{i*}_{n}= \psi^{i}_{n}\pm i \psi^{i+1}_{n}
\;,\;\; i=2,4,6,8 \;, \nonumber
\eeqa
with the commutation relations $[\alpha_m,\beta_{-n}] = -2 \delta_{mn}$, 
$[\beta_m^i,\beta_{-n}^{i*}] = 2 \delta_{mn}$, 
$\{\chi^A_m,\chi^B_{-n}\} = -2 \delta_{mn}$, 
$\{\chi_m^i,\chi_n^{i*}\} = 2 \delta_{mn}$.
For the RR zero modes, which satisfy a Clifford algebra and are thus 
proportional to $\Gamma$-matrices, $\psi^\mu_o = i \Gamma^\mu /\sqrt{2}$,
$\tilde \psi^\mu_o = i \tilde \Gamma^\mu /\sqrt{2}$,
one can construct similarly the creation-annihilation operators
$$
a,a^* = \frac 12 (\Gamma^0 \pm \Gamma^1) \;,\;\; 
b^i,b^{i*} = \frac 12 (-i\Gamma^i \pm \Gamma^{i+1}) \;,
$$
satisfying the usual algebra
$\{a,a^*\} = \{b^i,b^{i*}\} = 1$
(and similarly for tilded operators). 
In this way, any rotation or boost will reduce to a simple phase 
transformation on the modes. In fact, for an orbifold rotation ($g_a = 
e^{2\pi i z_a}$) one has
\beqa
\label{orb}
\al \beta _n^a \rightarrow g_a \beta_n^a \;,\;\;
\chi_n^a \rightarrow g_a \chi_n^a \;,\;\;
b^a \rightarrow g_a b^a \;, \nn \\
\al \beta_n^{a*} \rightarrow g^*_a \beta_n^{a*} \;,\;\;
\chi_n^{a*} \rightarrow g^*_a \chi_n^{a*} \;,\;\;
b^{a*} \rightarrow g^*_a b^{a*} \;.
\eeqa
whereas for a boost of rapidity $v$,
\beqa
\label{boost}
\al \alpha_n \rightarrow e^{-v} \alpha_n \;,\;\;
\chi^A_n \rightarrow e^{-v} \chi_n^A \;,\;\;
a \rightarrow e^{-v} a \;, \nn \\
\al \beta_n \rightarrow e^{v} \beta_n \;,\;\;
\chi_n^B \rightarrow e^{v} \chi_n^B \;,\;\;
a^* \rightarrow e^{v} a^* \;.
\eeqa

The boundary state which solves the boundary conditions
 can be factorized into a bosonic 
and a fermionic part; the latter can be further split into zero mode and 
oscillator parts, and finally
$$ 
|B> = |B>_B \otimes |B_{o}>_F \otimes |B_{osc}>_F \;.
$$

Let us now look at the internal coordinates.
An orbifold compactification can be obtained by identifying points in the 
compact part of space-time which are connected by discrete rotations 
$g = e^{2\pi i \sum_a z_a J_{aa+1}}$ on some of the compact pairs 
$X^a$,$\chi^a$, $a=4,6,8$. 
In order to preserve at least one supersymmetry, one 
has to impose $\sum_a z_a = 0$.

Three cases can be considered: 
toroidal compactification on $T_6$ ($N=8$ SUSY, 
$z_4=z_6 = z_8 = 0$) and orbifold compactification on $T_2 \otimes T_4/Z_2$ 
($N=4$ SUSY, $z_4= -z_6 = \frac 12$, $z_8 = 0$) and $T_6/Z_3$ ($N=2$ SUSY,
$z_4,z_6 = \frac 13, \frac 23$, $z_8 = -z_4 - z_6$).

The spectrum of the theory now contains additional twisted sectors, in which 
periodicity is achieved only up to an element of the quotient group $Z_N$. 
These twisted states exist at fixed points of the orbifold, and can thus occur
only for 0-branes localized at one of the fixed points. We will not discuss
this case here (see \cite{Hins1}).  

Finally, in all sectors, one has to project onto invariant states to get the 
physical spectrum of the theory which is invariant under orbifold rotations. 
In particular, the physical boundary state is given by the projection
$|B_{phys}>= \sum_k |B,g^k>/N$, in terms of the twisted boundary states 
$|B,g^k> = g^k|B>$.


Let us now concentrate in
 a particular 3-brane configuration. In the static case, 
we take Neumann boundary
conditions for time, Dirichlet 
 for space and mixed for 
each pair of compact directions, say Neumann for the $a$ directions and 
Dirichlet for the $a+1$ directions.

The boundary state has to satisfy in  the compact directions the
following conditions
\beqa
\al (\beta^a_n + \tilde\beta^{a*}_{-n} )|B>_B=0 \;,\;\;
(\beta^{a*}_n + \tilde\beta^{a}_{-n} )|B>_B=0 \;, \nn \\
\al (\chi^a_n +i\eta \tilde\chi^{a*}_{-n} )|B_{osc},\eta>_F=0 \;,\;\;
(\chi^{a*}_n +i\eta \tilde\chi^{a}_{-n} )|B_{osc},\eta>_F=0 \;, \nn \\
\al (b^a +i\eta \tilde b^{a*})|B_{o},\eta>_F=0 \;,\;\;
(b^{a*} +i\eta \tilde b^{a})|B_{o},\eta>_F=0 \;. \nn
\eeqa
We define the spinor vacuum $|0> \otimes |\tilde{0}>$ such that
$b^a|0> = \tilde b^{a} |\tilde{0}> = 0$. 
However, the 
boundary state is not invariant under orbifold rotations, under which the 
modes of the fields transform as in eq. (\ref{orb}) and the  spinor 
vacuum as $|0> \otimes |\tilde{0}> \rightarrow g_a |0> \otimes |\tilde{0}>$.
This was expected since a $Z_N$ rotation mixes two directions with 
different boundary conditions, 
and thus the corresponding closed string state does not need 
to be invariant under $Z_N$ rotations.
One finds for the compact part of the twisted boundary state
\beqa
\label{bs3}
\al |B,V,g_a>_B= \exp \left\{-\frac{1}{2}\sum_{n > 0}
(g_a^2 \beta^a_{-n}\tilde\beta^{a}_{-n} + 
g_a^{*2}\beta^{a*}_{-n}\tilde\beta^{a*}_{-n})\right\}|0> \;, \nn \\
\al |B_{osc},V,g_a,\eta>_F=\exp \left\{\frac{i\eta}{2}\sum_{n > 0}
(g_a^{2} \chi^a_{-n}\tilde\chi^{a}_{-n} 
+ g_a^{*2} \chi^{a*}_{-n}\tilde\chi^{a*}_{-n})\right\}|0> \;, \\
\al |B_o,V,g_a,\eta>_{RR} = g_a \exp \left\{-i\eta g_a^{*2} b^{a*} 
\tilde b^{a*} \right\}|0> \otimes |\tilde{0}> \;. \nn
\eeqa

After the GSO projection, the total partition functions for a given relative
angle $w_a$ turn out to be
\beqa
\al Z_B=16 i \sinh v q^{\frac 13} f(q^2)^4 
\frac 1{\vartheta_1(i \frac v\pi|2il)} 
\prod_a \frac {\sin \pi w_a}{\vartheta_1(w_a|2il)} \;, \\
\al Z_F=q^{-\frac 13}f(q^2)^{-4}
\left\{\vartheta_2(i\frac{v}{\pi}|2il)\prod_a \vartheta_2(w_a|2il) 
\right. \nn \\ \al \qquad \qquad \qquad \qquad \; \left.
-\vartheta_3(i\frac{v}{\pi}|2il)\prod_a \vartheta_3(w_a|2il)
+\vartheta_4(i\frac{v}{\pi}|2il)\prod_a \vartheta_4(w_a|2il)\right\} \label{pf}
 \\
\al  \quad \;\; \sim \left\{
\begin{array}{l}
V^4 \;\;,\;\;w_a=0 \\
V^2 \;\;,\;\;w_a \neq 0 
\end{array}
\right. \;.
\eeqa
Recall that to obtain the invariant amplitude, one has to average over all 
possible angles $w_a$.

In the large distance limit $l \rightarrow \infty$, explicit results with
exact dependence on the rapidity can be obtained 
from the above expression and compared to 
a field theory computation. One finds the following behaviors, according
to the compactification scheme:


\beqa
\al {\cal A}(w_a) \sim 4 \prod_a \cos \pi w_a \cosh v - \cosh 2v - 
\sum_a \cos 2 \pi w_a \;, \nn \\
\al {\cal A} \sim \left\{
\begin{array}{l}
4 \cosh v - \cosh 2v - 3 \sim V^4\;\;,\;\; T_2 \otimes T_4/Z_2 \;,\; T_6\\
\cosh v - \cosh 2v \sim V^2 \;\;,\;\; T_6/Z_3 \label{lipb}
\end{array}
\right. \;.
\eeqa

Let us now compare the large distance interactions of the two
moving branes found from the string formalism with the
field theory results. At large distances we look for the
Feynman graphs representing the exchange of massless particles
which can be either a scalar,  a vector or a 
graviton. 
Since we consider two branes of the same nature 
the scalar and the graviton give attraction  while
the vector gives repulsion. 

The net result in the static case is zero, since
the branes are BPS states, and this is what is obtained from
the Riemann identity in the string formalism \cite{Polch}. But when the
velocity is different from zero, the various contributions
are unbalanced. By comparing the velocity dependence with
what is obtained from Feynman graphs one can tell which kind of
particles are actually coupled to the branes, in the various
compactifications.

 We treat the branes as spinless particles of mass 
and charge equal to 1.
The exchange of a scalar gives then
\beq
{\cal S} = \frac{1}{k_{\perp}^2}
\eeq
where $k$ is the momentum transfer between the two branes.
In the so-called eikonal approximation in which the branes go 
straight (which is the standard setting for
describing the branes' interaction at nonsmall distances),
$k$ has only space components and is orthogonal to $\vec V$.

The vector is coupled to the current, which in the eikonal
approximation is proportional to the momentum, 
$J^{\mu}(V)\equiv(cosh(v), sinh(v))$. Note that
$J^{\mu}k_{\mu}=0$. Taking one of the branes at rest, the vector exchange
is
\beq
{\cal V}= J^{\mu}(V)J_{\mu}(0) \frac{1}{k_{\perp}^2}=
       -\frac{cosh(v)}{k_{\perp}^2}
\eeq

The graviton is coupled to the brane's energy-momentum tensor
$T^{\mu\nu}=J^{\mu}J^{\nu}$. Therefore the graviton exchange  in $d$-dimensions
is
\beq
{\cal G}= 2(T^{\mu\nu}(V)
        -\frac{\eta^{\mu\nu}}{d-2}T^{\rho\sigma}(V)\eta_{\rho\sigma})
       T_{\mu\nu}(0) \frac{1}{k_{\perp}^2}
     = \frac{cosh(2v)+\frac{d-4}{d-2}}{k_{\perp}^2}\,.   
\eeq
Thus the nature of the
various contributions to the branes' interaction 
can be read from the rapidity
dependence of the $l \to \infty$ limit of the amplitude (\ref{pf}),
and is the following for $d=4$
\beqa
4 \cosh v - \cosh 2v - 3 \quad \a \Leftrightarrow \a \quad
\mbox{$N=8$ Grav. multiplet} \;, \nn \\
\cosh v - \cosh 2v \quad \a \Leftrightarrow \a \quad
\mbox{$N=2$ Grav. multiplet} \;
\eeqa
        
\noindent In the second case, the two branes interact through the exchange
of the RR vector and the universal graviton with no scalar exchange. In 
terms of the N = 2 SUSY theory these systems couple only to
the graviton and its N = 2 partner, the graviphoton. 
>From the pattern
of cancellation \cite{pollard} these branes seem to correspond to
classical extremal Reissner-Nordstr\"{o}m blackholes. 
We present the evidence to support this conjecture in the next section.

\section{N=2 black hole supergravity solutions}

 BPS saturated solutions of four
dimensional N=2 supergravity coupled to N=2 vector multiplets have been
discussed in many recent papers. The simplest class of solutions is given
by the double extremal N=2 black holes with non vanishing electric and
magnetic charges. For this type of solution the values of the scalar moduli
fields which follow from a minimization of the N=2 central charge,
take constant values over the entire spacetime. In more general cases
of non constant moduli, the internal space does not decouple from the
four dimensional spacetime. In particular in static extremal N=2 black hole 
solutions the vector multiplet moduli
vary over the uncompactified space 
and one can argue that special or singular points
in the internal space are related to special or singular points in
spacetime (like horizons or curvature singularities).

The  concept of
{\it double-extremal} black hole was introduced in reference \cite{kal}.
 Non-extremal black holes have two horizons.
When they coincide the black hole is called extremal. As solutions of
supergravities, the mass of the extremal black hole depends on moduli
as well as on quantized charges. 
Double-extremal
black holes are extremal, supersymmetric black holes with the extremal value
of the ADM mass equal to the Bertotti Robinson mass.
They have constant moduli both for vector multiplets
as well as for hypermultiplets but the electric and magnetic
charges in each gauge group are unconstrained.       
We will obtain a four dimensional double extremal
black hole by compactifying an exact solution of type IIB supergravity
in 10 dimensions on a 3-cycle of the generic threefold ${\cal M}_3^{CY}$.

Let us start by considering the field equations of Type IIB supergravity
in 10 dimensions, namely

\begin{eqnarray}
\label{E10}
R_{MN} \!\!\!\! &=& \!\!\!\! T_{MN} \\
\nabla _M F_{(5)}^{MABCD}=0 \,
\; & \longleftarrow & \;  F^{(5)}_{G_1 \dots G_5}
= \frac{1}{5!} \, \epsilon_{G_1 \dots G_5 H_1 \dots H_5}\,F_{(5)}^{H_1 \dots H_5}
\end{eqnarray}
where $T_{MN}= 1/(2 \cdot 4!) \, F^{(5)}_{M \cdot \cdot \cdot\cdot } \,
F^{(5)}_{N \cdot \cdot \cdot \cdot }$ is the traceless energy--momentum tensor
of the R--R 4--form $A_{(4)}$ to which the 3--brane couples and
$F_{(5)}$   the corresponding self--dual field strength.
The tracelessness of $T_{MN}$ 
and the absence of couplings to the dilaton
(see for instance \cite{bac}), allows for zero curvature 
solutions in ten dimensions. 

For the metric we make a block--diagonal ansatz.
We  take for the four dimensional part $g^{(4)}_{\mu\nu}$ 
the extremal R-N black
hole solution
which depends only on the corresponding non--compact coordinates $x^\mu$. 
The Ricci--flat compact part
depends only on the internal coordinates $y^a$ (this corresponds to
choosing the unique Ricci flat
K\"ahler metric on ${\cal M}^{CY}_3$)

\begin{equation}
ds^2 = g^{(4)}_{\mu\nu}(x)dx^\mu dx^\nu + g^{(6)}_{ab}(y) dy^a dy^b
\end{equation}

In general, the compact components of the metric depend on the non--compact
coordinates
$x^\mu$ through the moduli which parametrize the deformations of the Kahler
class or the complex structure.
In Type IIB compactifications such moduli belong to
hypermultiplets and vector multiplets. In our case, however, where the
Hodge number $h^{(2,1)}=0$,
there are no vector multiplet scalars that would couple non--minimally to
the gauge fields 
and the hypermultiplet scalars can be set  to zero since they do not
couple  to the unique gauge field, namely the graviphoton
(therefore $g_{ab}(x,y) = g_{ab}(y)$).

The 5--form field strength can be generically decomposed in the basis of all
the harmonic 3--forms of the CY manifold $\Omega^{(i,j)}$
\begin{equation}
\label{F5}
F_{(5)}(x,y)=F^0_{(2)}(x) \wedge \Omega^{(3,0)}(y) +
\sum_{k=1}^{h^{(2,1)}} F^k_{(2)}(x) \wedge \Omega_k^{(2,1)}(y) + \mbox{c.c.}
\end{equation}
In the case at hand, however, only the graviphoton $F^0_{(2)}$ appears in the
general ansatz (\ref{F5}), without any additional vector multiplet field
strength $F_{(2)}^k$. We conveniently normalize
\begin{equation}
\label{F5noi}
F_{(5)}(x,y)= \frac 1{\sqrt{2}} F^0_{(2)}(x) \wedge \left(\Omega^{(3,0)} +
\bar \Omega^{(0,3)}\right)
\end{equation}
Notice that this same ansatz is  consistent  for any double--extremal
solution even for a more generic CY.

With these ans\"atze, eq. (\ref{E10}) reduces to the usual four--dimensional
Einstein equation with a graviphoton source. The compact part is
identically satisfied. 
The four--dimensional Lagrangian is obtained by carrying out the explicit
integration over the CY. Choosing 
an appropriate normalization for $\Omega^{(3,0)}$ and
$\bar \Omega^{(0,3)}$ such that $\left\|\Omega^{(3,0)}\right\|^2=V^2_{D3}/V_{CY}$
(since the volume of the corresponding 3--cycle is precisely the volume $V_{D3}$
of the wrapped 3--brane) one has
($z^a = 1/\sqrt{2} (y^a + i y^{a+1})$ and $d^6y = i d^3z d^3\bar z$)
\begin{equation}
\label{id}
\int_{CY} d^6y \sqrt{g_{(6)}} =  V_{CY} \;,\;\; 
i \int_{CY} \Omega^{(3,0)} \wedge \bar \Omega^{(0,3)} =  
V^2_{D3} = \int_{CY} d^6y \sqrt{g_{(6)}} 
\left\|\Omega^{(3,0)}\right\|^2
\end{equation}
and then
\begin{equation}
{\cal S}=\frac 1{2 \kappa_{(4)}^2}\int d^{4}x \sqrt{g_{(4)}}\left(R_{(4)} -
\frac 1{2 \cdot 2!} \mbox{Im}{\cal N}_{00} F^0_{\mu \nu}F^{0|\mu \nu} \right)
\end{equation}
where $\kappa_{(4)}^2 = \kappa_{(10)}^2 / V_{CY}$ and 
$\mbox{Im}{\cal N}_{00}= V_{D3}^2/V_{CY}$. 
In the general case (eq. (\ref{F5})) integration over the CY gives rise
to a gauge field kinetic term of the standard form: 
$\mbox{Im} {\cal N}_{\Lambda \Sigma} F^{\Lambda} F^{\Sigma}+\mbox{Re} 
{\cal N}_{\Lambda \Sigma}F^{\Lambda} {^*F}^{\Sigma}$, where 
$\Lambda,\Sigma=0,1,...,h^{(1,2)}$.
As well known (from now on $F_{(2)}^0\equiv F$), the four--dimensional Maxwell--Einstein 
equations of motion following from this Lagrangian admit the extremal R--N black hole 
solution (in coordinates in which the horizon is located at $r=0$)
\begin{equation}
\label{RNg}
\begin{array}{cclcccl}
g_{00} \!\!\!&=&\!\!\! 
\displaystyle{-\left(1 + \frac {\kappa_{(4)} M}r \right)^{-2}} &,\;&
g_{mm} \!\!\!&=&\!\!\!
\displaystyle{\left(1 + \frac {\kappa_{(4)} M}r \right)^2} \\ [3mm]
F_{m0} \!\!\!&=&\!\!\! \displaystyle{\kappa_{(4)} \, e_{0} \, \frac {x^m}{r^3} 
\left(1 + \frac {\kappa_{(4)} M}r \right)^{-2}} &,\;& 
F_{mn} \!\!\!&=&\!\!\! 
\displaystyle{\kappa_{(4)} \, g_{0} \, \epsilon_{mnp} \frac {x^p}{r^3}}
\end{array}
\end{equation}
where $m,n,p=1,2,3$. The extremality condition is $M^2 = (e^2 + g^2)/4$, where for 
later convenience we parametrize the solution with
\begin{equation}
\label{char}
M=\frac {\hat \mu}4 \;,\;\; e = e_{0} \sqrt{\frac {V_{D3}^2}{V_{CY}}}=
\frac {\hat \mu}2 \cos \alpha \;,\;\; 
g = g_{0}  \sqrt{\frac {V_{D3}^2}{V_{CY}}}=\frac {\hat \mu}2 \sin \alpha
\end{equation}
The parameter $\hat \mu$ is related to the 3--brane tension $\mu$ through
$\hat \mu = \sqrt{V_{D3}^2/V_{CY}} \mu$, and the angle $\alpha$ 
depends on
the way the 3--brane is wrapped on the CY. Notice that the charges 
with respect to 
the gauge field $A^\mu$ are $e_0$ and $g_0$, but since the kinetic term, and 
correspondingly the propagator of $A^\mu$, is not canonically normalized, the 
effective couplings appearing in a scattering amplitude are rather $e$ and 
$g$, 
which indeed satisfy the usual BPS condition. 
Further, at the quantum level, $e$ and $g$ are quantized as a consequence 
of Dirac's
condition $e g = 2 \pi n$; correspondingly, the angle $\alpha$ can take only 
discrete
values and this turns out to be automatically implemented in the 
compactification \cite{bis}.

This ends the field theory side of the computation. Let us now compare
with the microscopic 
string theory description of the same black--hole introduced in the
previous section.

The interaction between two D3-branes compactified on $T^6/Z_3$ in relative
motion, eqs. (\ref{pf}) and (\ref{lipb})
for large impact parameters, can be rewritten as

\begin{equation}
\label{amp}
{\cal A} = \frac {\hat \mu^2}{4} \left(\cosh v - \cosh 2v \right) 
\int dt \Delta_3(r)
\end{equation}
where $\Delta_3(r)$ is the 
three--dimensional Green function, $r = \sqrt{b^2 + \sinh^2 v t^2}$ and
$\vec b$ is the impact parameter.
This four--dimensional 
configuration comes from the following effective action 
\begin{equation}
{\cal S}=\int d^{4}x \sqrt{g}\left(R -
\frac 12 \left(\partial \phi \right)^2 -\frac 1{2 \cdot 2!} 
 e^{-a \phi}F_{(2)}^2 \right)
\end{equation}
where $a=0$ for the R--N black hole and $a\ne 0$ for the $0$-brane. We
concentrate in the first case for which 
the general 
electric extremal solution of this Lagrangian is \cite{stelle}
\begin{equation}
ds^2 = - H(r)^{-2} dt^2 + H(r)^{2} d \vec x \cdot d \vec x \;,\;\;
\phi = 0 \;,\;\;
A_0 = 2 \, H(r)^{-1}
\end{equation}
where
 $H(r)$ satisfies the three-dimensional Laplace equation and can be taken 
to be of the form $H(r) = 1 + k \Delta_3(r)$.
The relevant asymptotic long range fields are thus
$$
h_{00} = 2 \, k \, \Delta_3(r) \;,\;\; A_0 = 2 \, k \, \Delta_3(r)
$$
Comparing with eq. (\ref{amp})
we find that the R--N solution corresponds to  $k=\hat \mu / 4$.

An equivalent way of 
analyzing this configuration and providing more elements to identify the 
D3-brane with the general 
R--N $\times$ CY solution discussed before, is to compute one--point functions 
$\langle\Psi\rangle=\langle\Psi|B\rangle$ of the massless fields of 
supergravity and compare them with the linearized long range fields of the 
supergravity R--N black hole solution (\ref{RNg}).. This second method 
presents the
advantage of yielding direct information on the coulpings with the 
massless fields
of the low energy theory.

Let us consider the case in which the internal directions of the
D3-brane form an
arbitrary common angle
$\theta_0$ with the $X^{a}$ directions in
each of the 3 planes 
$X^a,X^{a+1}$ (actually, we could have chosen 3 different angles
in the 3 planes, but only their sum will be relevant).
The $Z_3$ projection is implemented by 
$|B\rangle = \frac 13 \sum_{\{\Delta \theta\}}
|B_3(\theta = \Delta \theta + \theta_0)\rangle$,
where the sum is over $\Delta \theta = 0, 2\pi/3,4\pi/3$.
It is obvious form this formula that $|B\rangle$ is a periodic function of the
parameter $\theta_0$ with period $2\pi/3$. Therefore, the physically distinct
values of $\theta_0$ are in $[0,2\pi/3]$ and define a one parameter family of
$Z_3$--invariant boundary states, corresponding to all the possible harmonic 
3--forms on $T^6/ Z_3$, as we will see.
Notice that requiring a fixed finite volume $V_{D3}$ for the 3--cycle on which
the D3--brane is wrapped implies discrete values for $\theta_0$ \cite{bis}.
The compactification process restricts the
momenta entering the Fourier decomposition of $|B\rangle$ to belong to the
momentum lattice of $T^6/ Z_3$. Since the massless supergraviton
states $|\Psi\rangle$ carry only space time momentum, the compact part of
the boundary state will contribute a volume factor which turns the 
ten--dimensional D3--brane tension $\mu=\sqrt{2\pi}$ into the 
four--dimensional 
black hole charge $\hat \mu = \sqrt{V_{D3}^2/V_{CY}} \mu$ \cite{bis}, and some 
trigonometric functions of $\theta_0$ to be discussed below.

Using the technique of ref. \cite{torino}, the relevant one--point functions
on $|B_3(\theta)\rangle$ for the graviton and 4--form states $|h\rangle$ and
$|A\rangle$ can be computed and one finds, by comparing with the
boundary state result, that the
 electric and magnetic charges are
\begin{equation}
\label{chars}
e = \frac 
{\hat \mu}2 \cos 3 \theta_0 \;,\;\; g = \frac {\hat \mu}2 \sin 3 \theta_0
\end{equation}
 Comparing with eq. (\ref{char}) 
one obtains
$\alpha = 3 \theta_0$ 
and therefore the ratio between $e$ and $g$ depends on the 
choice of the 3--cycle, as anticipated. Also, as explained, only discrete 
values of
$\theta_0$ naturally emerge requiring a finite volume. 

Further evidence for the identifications (\ref{chars})
  comes from the computation
of the electromagnetic phase--shift 
between two of these configurations with different
$\theta_0$'s, call them $\theta_{1,2}$.
Since the four--dimensional electric and magnetic charges of the
two black holes are then different, there should be both an even and an odd
contribution to the phase--shift coming from the corresponding R--R spin structures.
Indeed, one correctly finds \cite{bis}
\begin{equation}
{\cal A}_{even} \sim \frac {{\hat \mu}^2}4 
\cos 3 \left(\theta_{1} - \theta_{2} \right) =
e_1 e_2 + g_1 g_2 \;,\;\; {\cal A}_{odd} \sim 
\frac {{\hat \mu}^2}4 \sin 3 \left(\theta_{1} - \theta_{2} \right) =
e_1 g_2 - g_1 e_2
\end{equation}

Therefore  the asymptotic gravitational and electromagnetic fields
of the R--N black hole, eqs. (\ref{RNg}) are correctly reproduced.
This confirms that our boundary state describes a $D3$--brane
wrapped on $T^6/ Z_3$, falling in the class of regular four--dimensional
R--N double--extremal black holes obtained by wrapping the self--dual
D3--brane on a generic CY threefold. This boundary state encodes the
leading order couplings to the massless fields of the theory, and allows the
direct determination of their long range components, falling off like $1/r$ in
four dimensions. The subleading post--Newtonian corrections to these fields
arise instead as open string higher loop corrections, corresponding to string
world--sheets with more boundaries; 
from a classical field theory point of view, 
this is the standard replica of the source in the tree-level 
perturbative evaluation 
of a non--linear classical theory. 
\vskip 7pt
To conclude, let us comment that one could interpret  
the $Z_3$--invariant  boundary state as describing
the three D3--branes 
superposition at angles ($2 \pi/3$) in a $T^6$ compactification.
As illustrated in \cite{tow} such intersection preserves precisely $1/8$
supersymmetry, as a single D3--brane does on $T^6/ Z_3$. For toroidal
compactification this is not enough, of course, because at least 4
intersecting 
D3--branes are needed in order to get a regular solution \cite{bal}.

Finally, 
since this extremal R--N configuration is constructed by a single D3--brane, 
the question naturally arises
of understanding the microscopic origin of its entropy.

\nobreak

 \end{document}